\documentclass[12pt]{book}

\usepackage[dvips]{graphicx,color}
\usepackage{makeidx,physics,cosmology}

\makeauthorindex
\makeindex

\BookTitle{Frontier in Astroparticle Physics and Cosmology}
\CopyRight{\copyright 2004 by Universal Academy Press, Inc.}

\begin{document}

\BookTitle{\itshape Frontier in Astroparticle Physics and Cosmology}
\CopyRight{US-04-01
}
\pagenumbering{arabic}
\chapter{
Casimir Effect in the Brane World
}

\author{%
Shoichi ICHINOSE\\
{\it 
Laboratory of Physics, 
School of Food and Nutritional Sciences, 
University of Shizuoka, 
Yada 52-1, Shizuoka 422-8526, Japan}\\
Akihiro MURAYAMA\\
{\it 
Department of Physics, Faculty of Education, Shizuoka University,
Shizuoka 422-8529, Japan}}
%
%
\AuthorContents{S.\ Ichinose and A.\ Murayama} 

\AuthorIndex{Ichinose}{S.} 
\AuthorIndex{Murayama}{A.}

\section*{Abstract}
Taking the Mirabelli-Peskin model, we examine
the Casimir effect in the brane world.
It is compared with that of the ordinary
Kaluza-Klein theory.
\section*{}
{\bf 1}\ {\it Introduction}\quad
Casimir effect is a general phenomenon which takes place
in the quantum field theories with a boundary. 
The quantum fluctuation of the vacuum 
induces a potential depending on the global parameters
(separation length, depth, width, etc) of the boundary.
This is purely a boundary effect in the quantum system. 
The famous gravitational example 
is that of the Kaluza-Klein theory. Appelquist and
Chodos\cite{AC83} examined the 5D bulk quantum effect of the
system and found the following vacuum energy:
$
V_{eff}=\Lambda^5/8\pi+5\beta/(2l)^5{\phi_c}^{5/3}\ ,\ 
\beta=-0.394
\ \mbox{(1)}
$. 
The first term is the cosmolgical term which is
quintically divergent and depends on the cut-off $\Lambda$.
The physical meaning is still now obscure.
The second term, on the other hand, is finite
and cut-off independent. Furthermore it is
gauge-independent\cite{SI85}.
This term is considered meaningful and 
corresponds to the Casimir energy. 
The negative sign of the coefficient $\beta$ means
the Casimir force is {\it attractive}. The circle
of the extra coordinate space tends to shrink.
This fact supports our common image that
the higher dimentional models are compactified
to 4D space-time at the low energy 
(compared to the Planck energy) level.
Recently new models
(such as Ekpyrotic/Cyclic model\cite{KOST01PR,ST02PR})
based on the brane configuration appear and are examined. 
In these models the bulk quantum
fluctuation generally induces the Casimir energy. We examine
it using a typical model (based on the $S^1/Z_2$ orbifold) of the brane world.

{\bf 2}\ {\it Mirabelli-Peskin Model\cite{MP97} and Background Fields}\quad
We take a 5D bulk theory ${\cal L}_{bulk}$ which is
coupled with a 4D matter theory ${\cal L}_{bnd}$ on one "wall" at ${x^5}=0$
and with ${\cal L}'_{bnd}$ on the other "wall" at ${x^5}=l$:\ 
$
S=\int d^5x\{{\cal L}_{blk}+\delta(x^5){\cal L}_{bnd}+\delta(x^5-l){{\cal L}'}_{bnd}
\}
\ \mbox{(2)}
$. 
We consider the case:\ 
$
{\cal L}'_{bnd}=-{\cal L}_{bnd}
$. 
The bulk dynamics is given by the 5D super YM theory
($A^M, \Phi, \lambda^i, X^a$). 
We introduce a 4 dim chiral multiplet ($\phi, \psi, F$) on the walls. 
Using the ${\cal N}=1$ SUSY property of the bulk fields 
($A^m,\lambda_L,{\cal D}\equiv X^3-\nabla_5\Phi$),
we find the supersymmetric bulk-boundary coupling.
For the interest in the potential part, we focus on the
scalar fields part. The
following background fields are taken. For the boundary scalars,
$\eta=\mbox{const}, \eta^\dag=\mbox{const}$. For the
D-field on the boundary, $d=\mbox{const}$. As for the
bulk fields, $a_5$ and $\varphi$, 
we take the following forms 
which describe a localized (around $x^5=0$) configuration and
a natural generalization of the ordinary treatment of the vacuum.
$
a_{5\gamma}({x^5})={\bar a}_\gamma\,\epsilon ({x^5}),
\varphi_\gamma({x^5})={\bar \varphi}_\gamma\epsilon ({x^5})
\ \mbox{(3)}
\ ,$
where $\epsilon({x^5})$ is the {\it periodic sign function}
with the periodicity $2l$. ${\bar a}_\gamma$ and ${\bar \varphi}_\gamma$ are 
some positive constants. 
It is the {\it thin-wall limit} of a (periodic) kink solution
and shows the {\it localization} of the fields. 
The background fields, (3),
satisfy the required boundary condition and
the field equation   
for an appropriate
choice of ${\bar a}, {\bar \varphi}, \eta, \eta^\dag$ and $\chi^3$.

{\bf 3}\ {\it Effective Potential and Casimir Energy}\quad
The effective potential
is obtained from the eigenvalues of the "mass-matrix" derived
by "expanding" the Lagrangian (2) around the above
background fields. We examine the behaviour for two typical cases.

(A)$\eta=0,\eta^\dag=0$\ \ 
We look at the potential from the vanishing scalar-matter point.
In this case the {\it singular terms disappear} and 
the boundary quantum fluctuation and 
 the bulk one {\it decouple}. The former part gives, 
taking the {\it supersymmetric boundary condition}, 
the following potential:\ 
$
V^{eff}_{1-loop}
=\int\frac{d^4k}{(2\pi)^4}
\ln \{1-\frac{g^2}{4}\frac{d^2}{(k^2)^2}\}
\ \ 
\mbox{(4)}\ .
$ 
The form of (4) is similar to the 4D super QED\cite{Mil83NP}. 
We see the present model produces a desired effective potential on the brane.
The bulk part do {\it not} depend on the field $d$.
The eigenvalues depend only on 
the brane parameters, ${\bar a}$ and ${\bar \varphi}$, and the size of the extra space, $l$.
They give the scalar-loop contribution to the Casimir energy (potential).
We consider the large circle limit:\ 
${\hat g}^2\equiv \frac{g^2}{l}=\mbox{fixed}\ll 1,
{\hat a}=\sqrt{l}{\bar a}=\mbox{fixed},
{\hat \varphi}=\sqrt{l}{\bar \varphi}=\mbox{fixed},
l\rightarrow\infty$.
This is the situation where the circle is large
compared with the inverse of the domain wall height.
We notice, in this limit, all KK-modes equally contribute to the vaccum energy.
The eigenvalues of the bulk part of the "mass-matrix" can be
easily obtained. 
In particular, for the special case ${\hat a}=0$, the nontrivial
factor is only $k^2+{\hat g}^2{\hat \varphi}^2$. Hence each KK-mode
equally contributes to the vacuum energy as, 
$
V^{eff}_{1KK-mode}
\propto\int\frac{d^4k}{(2\pi)^4}
\ln \{1+{\hat g}^2\frac{{\hat \varphi}^2}{k^2}\}
\ \mbox{(5)}
$.
This quantity is quadratically divergent. 
After an appropriate normalization
the final form should become, based on the dimensional analysis, 
the following one.
$\frac{1}{l}V^{eff}_{Casimir}
={\hat g}^2(c_1\frac{{\hat \varphi}^2}{l^3}+c_2\frac{{\hat a}^2}{l^3})+O({\hat g}^4)
\ \ \mbox{(6)}
\ ,
$
where $c_1$ and $c_2$ are some finite constants which are calculable
after we know the bulk quantum dynamics sufficiently.
This is a {\it new} type Casimir energy.
Comparing the ordinary one (7) explained soon,
it is new in the following points:\ 
1) it depends on the brane parameters ${\hat \varphi}$ and ${\hat a}$
besides the extra-space size $l$;\ 
2) it depends on the gauge coupling ${\hat g}$;\ 
3) it is proportional to $1/l^3$.\ 
We expect the above result (6) are cancelled 
by the spinor and vector-loop contributions in the present SUSY theory.
The unstable Casimir potential do {\it not} appear in the SUSY theory.

(B)${\bar a}=0,{\bar \varphi}=0$\ \ 
In this case, 
we have no localized (brane) configuration. 
This is similar to the 5D Kaluza-Klein case mentioned
in the introduction. 
The eigenvalues for the bulk part
are commonly given by, 
$
\lambda_n=-k^2-({\frac{n\pi}{l}})^2\ ,\ n=1,2,3,\cdots
$. 
The eigenvalues are basically the same as those of the KK case \cite{AC83}.
They depend {\it only} on the radius parameter $l$.
It is the scalar-loop contribution to the Casimir potential. From the
dimensional analysis, after the renormalization, it has the form:\ 
$\frac{1}{l}V^{eff}_{Casimir}= \frac{\mbox{const}}{l^5}
\ \ \mbox{(7)}
\ .$
We expect again this contribution is cancelled 
by the quantum effect of the spinor and vector fields. 
The eigenvalues for the boundary part are obtained as 
a complicated expression involving the following terms:
$ S\equiv\eta^\dag\eta\ ,\ d^2=d_\alpha d_\alpha\ ,\ 
d\cdot V\equiv d_\alpha \, \eta^\dag T^\alpha\eta\ ,\ 
V^2\equiv (\eta^\dag T^\alpha\eta)^2 
\ \ \mbox{(8)}
$.
In the manipulation of eigenvalues search, 
we encounter the following comibination of terms:\ 
$
\delta(0)+\frac{1}{l}\sum_{m=1}^{\infty}
\frac{(\pi m/l)^2}{-\lambda-k^2-(\pi m/l)^2}
\ \mbox{(9)}
$. 
The first term comes from the {\it singular} terms
in (2), the second from the KK-mode sum.
Using the relation$\sum_{m\in{\bf Z}}1=2l\delta(0)$, 
the above sum leads to a {\it regular} quantity:
${\frac{1}{2}}\sqrt{\lambda+k^2}\coth\{l\sqrt{\lambda+k^2}\}$\ ($\lambda>-k^2$).
We have confirmed this "smoothing" phenomenon
occurs at the full 1-loop level.
In order to present an explicit form
we take an intersesting case:\ 
$d\cdot V\neq 0$, others=$0$. 
Eigenvalues come from the solutions of the
following equation:\ 
$
(\lambda+k^2)^2-\frac{g^3}{2}d\cdot V\frac{\sqrt{\lambda+k^2}}{2}
\coth l\sqrt{\lambda+k^2}=0
\ \mbox{(10)}
$.
Perturbatively (w.r.t. $g$),
two eigenvalues $\lambda_1,\lambda_2$ of the above equation satisfy\ 
$
\lambda_1\lambda_2=
(k^2)^2\left( 1-\frac{g^3}{4}d\cdot V
\frac{\sqrt{k^2}\coth l\sqrt{k^2}}{(k^2)^2}\right)
\ \mbox{(11)}
$.
The full-order eigenvalues, (10),
correspond to the 1-loop {\it full} effective potential.

{\bf 4}\ {\it Conclusion}\quad
We find a new type form
of the Casimir energy which is characteristic for
the brane model. When SUSY is broken in some mechanism,
the new type potential could become an important
distinguished quantity of the bulk-boundary system
from the ordinary KK system.


\end{document}